\documentstyle[aps,twocolumn,epsfig]{revtex}

\draft
\title{Quantum State Reconstruction Using Atom Optics}
\author{B.~T.~H.~Varcoe \thanks{Current address: Max-Planck-Institut f\"ur Quantenoptik, 85748 Garching, Germany},
R.~Sang, 
W.~R.~MacGillivray, and 
M.~C.~Standage
}
\address{
Griffith University,
4111 Nathan , Qld. Australia.\\
}

\date{Received \today}

\begin{document}

\maketitle

\abstract
We present a novel technique in which the total internal quantum state of an atom may be reconstructed via the measurement of the momentum transferred to an atom following its interaction with a near resonant travelling wave laser beam. We present the first such measurement and demonstrate the feasibility of the technique.
\endabstract


The development of methods for completely determining the quantum state of an atom or cavity field \cite{bodendorf} continues to be an important issue in atomic physics. For example, in the field of quantum optics, schemes have been recently proposed, based on the Stern-Gerlach effect, for completely determining the quantum state of an atom as a way of fully characterising the quantised electromagnetic field of a cavity mode \cite{walser}. The reconstruction of an internal atomic state is important for goals as far reaching as reading the end-state of a quantum gate operation \cite{wineland} and teleportation of internal atomic states \cite{cirac}. Walser et. al.\cite{walser} describe a scheme whereby the field state of a cavity could be transferred to the magnetic sublevels of the internal state of an atom by the technique of adiabatic passage, thus converting the problem to one of determining a quantum state of an atom. Atom deflection has been suggested as a probe of the photon number in a cavity \cite{freyberger} and atom deflection has also been used to measure the Mandel Q parameter of photon statistics for resonance fluorescence \cite{hoogerland,oldaker}. This technique provides a significant improvement over conventional photon counting schemes due to the poor efficiency of photon detection which affects measurements of the Q parameter. In the field of electron-atom collision physics, the measurement of the atomic collision density matrix using coincidence and superelastic scattering methods depends on the complete determination of the quantum state of the atom \cite{andersen2}.

The internal quantum state of an atom can be described by the density matrix formalism which contains all the information about the quantum system. In principle, if one can measure all elements of the density matrix, then complete reconstruction of the quantum state of a system is possible. Atomic states are relatively insensitive to decoherence and can maintain an initial quantum state for long time scales compared to the characteristic times for preparation and detection, and as such they are frequently used in quantum computing proposals where insensitivity to decoherence is a requirement \cite{briegel}.

In this Letter, a novel method is presented for completely determining the atomic density matrix of the internal state of an atom based on laser induced deflection of a state prepared sodium atom. The experiment is a realisation of a suggestion put forward by Summy et. al. \cite{summy3,summy,summy2} and represents the direct measurement of atomic density matrix elements for a state with angular momentum above J=1. The technique may also be used in the reconstruction of internal atomic metastable states. The motivation for this work was to develop an experimental tool that allows detailed information about the density matrix to be extracted from atoms. Using deflection with low laser powers, for which measurements are sensitive to the internal atomic structure of the atom. The technique relies on the difference in atomic absorption probabilities for individual optical hyperfine transitions that are not only dependent on the initial hyperfine populations, but also coherences between the hyperfine levels. Hence, atoms in different internal atomic states will be differentially deflected. It is shown that it is possible to reconstruct the internal state of the atom fully via the measurement of the centre of mass motion of an atom after it has interacted with well-defined polarised deflecting laser beams. Moreover, it is possible to restrict the number of measurements if only a small part of the density matrix is of interest. A measurement is presented whereby one of the features of the density matrix is probed, namely, the net angular momentum of the atomic state. The work presented here provides a ``proof in principle" of the deflection forces' sensitivity to the complete density matrix.

The F=2 ground state of a highly collimated sodium beam was prepared in a known density matrix\cite{varcoe}. With a suitable set of measurements of the deflected momentum we show that the electron charge cloud distribution can be completely reconstructed. The net angular momentum, or alternatively the charge cloud shape parameter $L_{\perp}$, was measured as it was varied from +2 to -2. 

\begin{figure}
\epsfxsize=8cm
\centerline{\epsffile{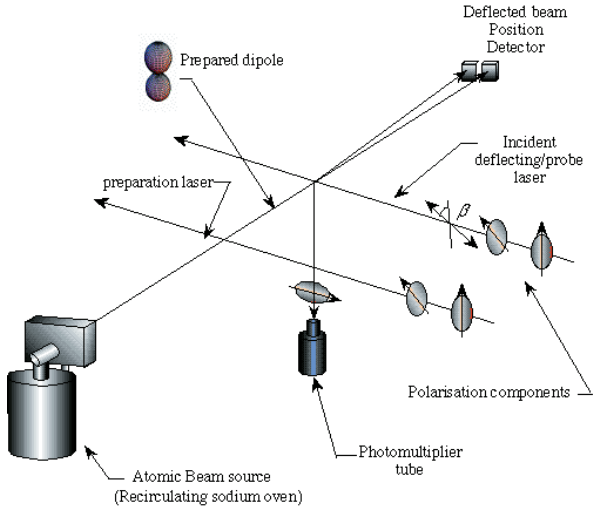}}
\vspace{20ex}
\caption{
The experimental setup. In the current experiment a preparation laser prepares an atomic orbital shape which is then deflected by the probe laser. A change in one aspect of the prepared dipole produces a corresponding change in the deflected position of the beam.
}
\label{setup}
\end{figure}

A schematic of the experimental setup is presented in Fig. \ref{setup}. A highly collimated, velocity selected, sodium beam passes through preparation and deflection laser interaction regions before travelling down a 1m flight tube to a scannable hot wire detector. Firstly the atoms interact with a travelling wave state preparation laser, with arbitrary but well defined elliptical polarisation. The atoms are deflected by a travelling wave laser with a pure polarisation ($\pi_{\rm 0,90}$, $\sigma^{\rm \pm}$). Both lasers will deflect the atomic beam providing an additional separation between the deflected and undeflected beams. The atomic beam is velocity selected using two chopper wheels and has a velocity width of about 10\%. The ellipticity of the preparation laser is defined by the angle of a quarter wave plate with respect to the $\pi_{\rm 0}$ linearly polarised preparation laser. The deflected distance was measured by fitting two Gaussian curves to the peaks due to the undeflected $F=1$ and deflected $F=2$ atom beams. 

For a laser with a polarisation mode $k$, incident on an atomic beam the deflection force is given by the diagonal elements of the atomic density matrix \cite{summy3,summy,summy2},

\begin{equation}
\vec{p}_{\rm k}=\sum_g D(t)^{\rm gg}_{\rm k}\rho_{\rm gg}^{\rm L}(t=0)
\label{mom}
\end{equation}
where $\rho_{\rm gg}^{\rm L}(t=0)$ represents the initial density matrix elements of the lower state of the laser transition in the reference frame $``L"$. The reference frame $``L"$ is defined by the deflection laser. The $z$-axis is defined by the laser polarisation and the $x$-axis is defined as the direction of the laser in the case of linear polarisation and in the direction of the atomic beam in the case of circular polarisation. The $D^{\rm gg}_{\rm k}(t)$ are deflection parameters relating to the optical pumping and momentum absorption process which occurs in an interaction time $t$. The preparation of the atomic state is defined in the ``natural frame" for which the co-ordinates are fixed \cite{andersen2}. The $z$-axis of the natural frame points in the direction of propagation of the laser and the $x$-axis in the direction of the atomic beam. Euler rotations achieve transformations of the density matrix between the two frames\cite{edmonds}. 

In this experiment differential deflection measurements are constructed using polarisations of the deflection laser. For example in analogy with Ref.\cite{andersen} an orientation parameter, $P_{\rm o}$, is defined,

\begin{equation}
P_{\rm o}=\vec{p}_{\rm \sigma^{\rm -}}-\vec{p}_{\rm \sigma^{\rm +}}
\label{exp_mom}
\end{equation}

This is called the orientation parameter by virtue of its definition in terms of natural frame density matrix elements;

\begin{equation}
P_{\rm o} = (D_{\rm \sigma^{\rm +}}^{\rm -2-2}(t)-D_{\rm \sigma^{\rm +}}^{\rm 22}(t))(\rho_{\rm 22}^{\rm N}-\rho_{\rm -2-2}^{\rm N})+D_{\rm \sigma^{\rm +}}^{\rm 11}\rho_{\rm 11}
\label{pc}
\end{equation}

Hence information about the density matrix can be extracted from a deflection measurement. The meaning of Eq. \ref{pc} becomes apparent when it is rewritten in terms of the natural frame charge cloud shape parameter $L_{\perp}$,

\begin{equation}
P_{\rm o} = K_{\rm 2}(t)L_{\perp}^{\rm {\pm}2}+K_{\rm 1}(t)L_{\perp}^{\rm {\pm}1}
\label{lperp}
\end{equation}

where $L_{\perp}^{\rm {\pm}2}$ refers to the angular momentum contained in the $m_{\rm F}=\pm 2$ states and the coefficients $K_{\rm 2}$ and $K_{\rm 1}$ are defined,

\begin{equation}
K_{\rm 2}(t)= \sigma_{\rm u}(D_{\rm \sigma^{\rm +}}^{\rm -2-2}(t)-D_{\rm \sigma^{\rm +}}^{\rm 22}(t)); \\
K_{\rm 1}(t)=D_{\rm \sigma^{\rm +}}^{\rm 11}.
\label{k}
\end{equation}

The $K_{\rm 1,2}(t)$ parameters contain all of the momentum exchange, optical pumping and normalisation information. These parameters can be calculated experimentally or theoretically using a simulation of the interaction \cite{summy,varcoe} and they are relatively insensitive to changes in experimental parameters \cite{mythesis}. 

\begin{figure}
\epsfxsize=8cm
\centerline{\epsffile{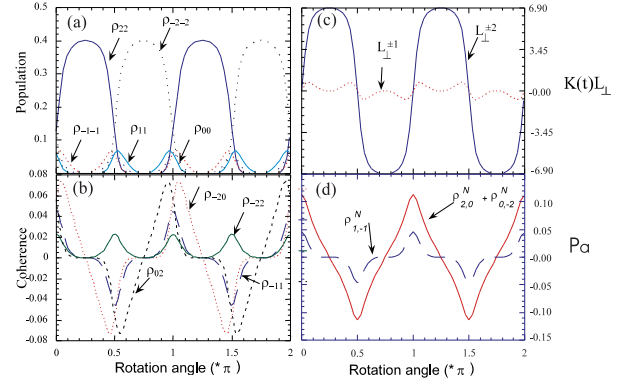}}
\vspace{20ex}
\caption{
Simulations of the prepared density matrix elements and the corresponding parameter variations. Fig. (a) and (b) presents the populations and coherences respectively that leave the preparation region. Fig. (c) and (d) presents the variation in the experimentally measureable parameters relating to the angular momentum ground states $m_{\rm F} = 0$, $m_{\rm F} = \pm 1$ and $m_{\rm F} = \pm 2$ and the coherences between these sublevels.  
}
\label{lperpfig}
\end{figure}

A simulation was performed by dividing the flight path of the atoms into 3 regions. The first was a QED simulation of the elliptically polarised preparation region followed by a free travel to the second region in which a QED simulation \cite{summy,varcoe} for a fixed polarisation was used to calculated the deflection of the atoms. Fig. \ref{lperpfig} presents the results of this simulation for experimental parameters that correspond to the values obtained in the experiment. Fig. \ref{lperpfig}a,b shows the populations and coherences between magnetic sublevels for atoms emerging from the preparation region. It can be seen that the populations have a maximum below 1, this is due to losses in the optical preparation stage to the F=1 ground state. 

Fig. \ref{lperpfig}c shows the actual momentum transferred to the atoms by the second interaction region in terms of the charge cloud shape parameters for a fixed $\sigma^-$ laser polarisation. The deflection parameters ($D_{\rm \sigma^{\rm \pm}}^{\rm mm}(t)$) and the optical pumping parameter ($K_{\rm 2}(t)$) have a constant value as a function of the preparation angle. For conditions approximating the experimental conditions, a value of $K_{\rm 2}(t)\approx3.45$ was obtained. Contributions to the deflection due to the $L_{\perp}^{\rm {\pm}1}$ component are small and can be neglected in the analysis of experimental results.

The reconstruction of the density matrix considered in this work requires the measurement of $9$ independent parameters, namely ($\rho_{\rm \pm2,\pm2}^{\rm N}$,$\rho_{\rm \pm2,\mp2}^{\rm N}$,$\rho_{\rm \pm1,\pm1}^{\rm N}$,$\rho_{\rm \pm1,\mp1}^{\rm N}$,$\rho_{\rm 0,0}^{\rm N}$) and a parameter relating to the total population of the state. Following the rules of the Euler rotations, it can be shown that combinations of the six available polarisations ($\pi_{\rm 0,90}$,$\pi_{\rm 45,135}$,$\sigma^{\rm \pm}$)and two different incident directions of the deflection laser allows all parameters to be obtained \cite{mythesis}. It is possible to determine individual matrix elements from both the diagonal and off diagonal parts of the matrix, for example a parameter $P_{\rm a}$ can be defined,

\begin{equation}
P_{\rm a} = (\vec{p}_{\rm 0}-\vec{p}_{\rm 90})+i(\vec{p}_{\rm 45}-\vec{p}_{\rm 135})
\label{Pa}
\end{equation}
where $\vec{p}_{\rm \theta}$ is the deflected momentum for a deflection laser polarisation angle $\theta$. In terms of the natural frame matrix elements Eq. \ref{Pa} becomes,

\begin{equation}
P_{\rm a} = K^{\rm a}_{\rm 2}(t)(\rho_{\rm 2,0}^{\rm N}+\rho_{\rm 0,-2}^{\rm N})+K^{\rm a}_{\rm 1}(t)(\rho_{\rm 1,-1}^{\rm N}) 
\label{Pa2}
\end{equation}
where 

\begin{equation}
K^{\rm a}_{\rm 2}(t)=2\sqrt{6}\sigma_{\rm u}((D_{\rm \pi}^{\rm 00}(t)-D_{\rm \pi}^{\rm 22}(t)); K^{\rm a}_{\rm 1}(t)=2D_{\rm \pi}^{\rm 11}(t)
\label{Ka1}
\end{equation}

In fact all density matrix elements can be measured by suitable arrangement of parameters. The coherence $\rho_{\rm 2,-2}^{\rm N}$ in Eq. \#\ref{Pg2} is particularly interesting as it can only be created by a multiple photon process. Such a coherence could be used to create macroscopic superpositions of photon numbers in a cavity, or alternatively the presence of this coherence could be used as an indication that the atom had interacted with a superposition of multiple photon states.   

\begin{equation}
P_{\rm g} = (\vec{p}_{\rm 0}+\vec{p}_{\rm 90})-(\vec{p}_{\rm 45}+\vec{p}_{\rm 135})
\label{Pg}
\end{equation}

\begin{equation}
P_{\rm g} = K^{\rm g}_{\rm 2}(t)\rho_{\rm 2,-2}^{\rm N}
\label{Pg2}
\end{equation}
where 

\begin{equation}
K^{\rm g}_{\rm 2}(t)=\sigma_{\rm u}((D_{\rm \pi}^{\rm 22}(t)+3D_{\rm \pi}^{\rm 00}(t)-4D_{\rm \pi}^{\rm 11}(t))
\label{Kg}
\end{equation}

The close resemblance of Eq. \#\ref{Pa} to Eq. \#\ref{Pg} is to be expected as the density matrix elements $\rho_{\rm 2,0}^{\rm N}$ and $\rho_{\rm 0,-2}^{\rm N}$ are directly related to the element $\rho_{\rm 2,-2}^{\rm N}$ \cite{andersen}.

\begin{figure}
\epsfxsize=8cm
\centerline{\epsffile{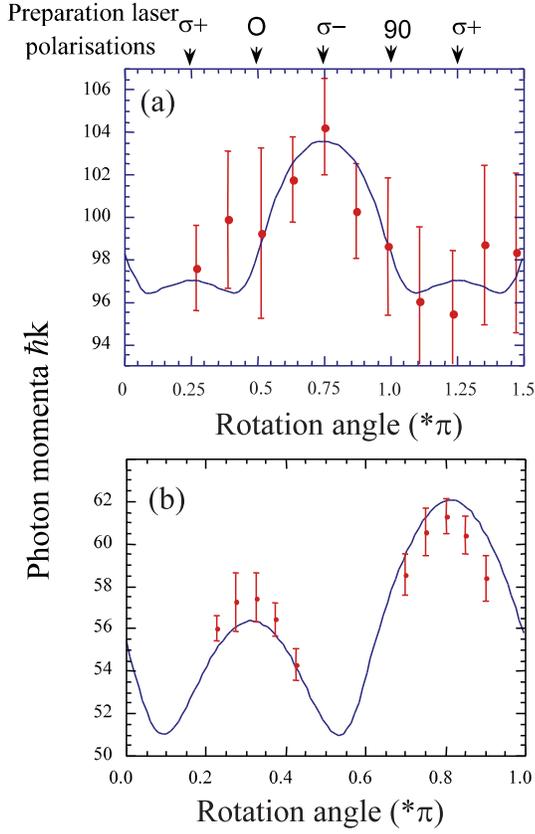}}
\vspace{20ex}
\caption{
Experimental measurements of deflected distances, measured in photon momenta, as a function of state preparation. Plot (a) presents an experimental scan as a function of the rotation angle of the laser polarisation along with the simulation of this deflection. Plot (b) presents a second experimental plot and associated simulation, taken for conditions that enhanced the visibility of the loss feature at $pi/4$.
}
\label{data}
\end{figure}

Fig. \ref{data} presents two measurements of the $P_{\rm o}$ parameter as the preparation laser polarisation was rotated through an angle of $2\pi$. The solid line is an evaluation of the simulation for experimental conditions of laser intensity and detuning. In Fig. \ref{data}a there is an additional feature in the deflection simulation at the angles $\pi/4$ and $5\pi/4$ that has not been resolved. This feature is a result of losses to the F=1 ground state in the preparation region. A second scan was taken for experimental conditions that maximised the loss and hence the feature. This is presented along with the associated simulation in Fig. \ref{data}b. 

\begin{figure}
\epsfxsize=8cm
\centerline{\epsffile{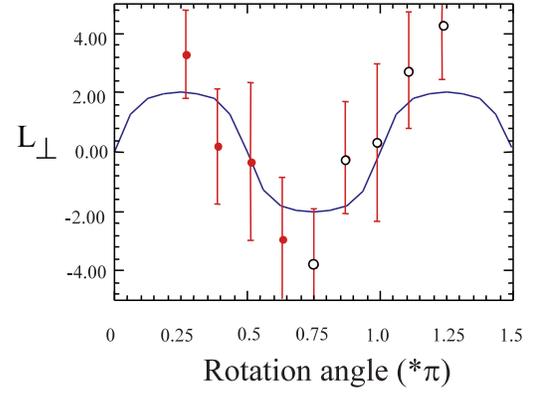}}
\vspace{20ex}
\caption{
The experimental evaluation of the parameter $L_{\rm \perp}^{\rm \pm 2}$, evaluated using Eq. \#2 as the prepared angular momentum state was varied between -2 and +2. The calculation is performed using the inherent symmetry of the data and is applied in two directions. The repeated evaluation is displayed with closed circles.
}
\label{result}
\end{figure}

From this deflection data, the parameter $L_{\rm \perp}^{\rm \pm 2}$ can be extracted. The preparation method has an inherent symmetry in that both $\sigma+$ and $\sigma-$ polarisations are contained in the elliptical polarisation. This symmetry has been used along with Eq. {\#}2 to construct Fig. \ref{result}, the symmetry can be applied in two directions and the repeated result is presented with closed circles. 

These results show that deflection measurements are sensitive to the polarisation of the preparation laser and hence to the atomic charge cloud shape. A complete measurement of the deflections due to a higher number of polarisations can be used to reconstruct the quantum state. 

We have observed the sensitivity of the deflection to an atomic charge cloud shape parameter and we have shown that a suitable set of measurements can be used to completely reconstruct a density matrix of an internal atomic state. The simulation demonstrates that the deflection is also a sensitive detector of ground state coherences. Work is currently underway to replace the sodium beam with a Zeeman cooled metastable Neon beam. This will have the effect of greatly increasing the resolution of the system by considerably reducing the losses owing to the absence of a distant ground state a cooled beam will also increase the perpendicular deflections of the beam. These effects combine to reduce the errors by two orders of magnitude.

\newpage

\newpage

\newpage

\newpage
\end{document}